\documentclass[reprint,
 amsmath,amssymb,
 aps,
]{revtex4-2}

\usepackage{graphicx}% Include figure files
\usepackage{dcolumn}% Align table columns on decimal point
\usepackage{bm}% bold math
\usepackage{xcolor}
\usepackage{mathptmx}
\usepackage{graphicx}% Include figure files
\usepackage{dcolumn}% Align table columns on decimal point
 
\usepackage{bm}% bold math
\usepackage{amssymb}
\usepackage{comment}
\usepackage{hyperref}
\begin{document}

\title{Long-distance axial spectral encoding using space-time wave packets}

\author{Layton A. Hall$^{1,2}$}

\author{Murat Yessenov$^{1,3}$}

\author{Miguel A. Romer$^{1}$}

\author{Ayman F. Abouraddy$^{1,*}$}

%\email{raddy@creol.ucf.edu}
\affiliation{$^{1}$CREOL, The College of Optics \& Photonics, University of Central~Florida, Orlando, FL 32816, USA}
\affiliation{$^{2}$Current address: Materials Physics and Applications - Quantum Division, Los Alamos National Laboratory, Los Alamos, NM 87545, USA}
\affiliation{$^{3}$Current address: Harvard John A. Paulson School of Engineering and Applied Sciences, Harvard University, Cambridge, MA, USA}
\affiliation{$^*$Corresponding author: raddy@creol.ucf.edu}

\begin{abstract}
Space-time wave packets (STWPs) are pulsed optical beams whose spatiotemporal structure enables propagation invariance. However, STWPs allow for a unique propagation configuration that we call axial spectral encoding, in which the spectrum on the propagation axis changes with distance. We demonstrate here axial spectral encoding over distances extending for hundreds of meters in an open-field laser range. We verify two distinct configurations: in the first, the on-axis spectrum blue-shifts or red-shifts with propagation distance; and in the second, the spectrum changes at a fixed axial position by altering an internal parameter of the STWP without modifying the wave packet spectrum. These results indicates the potential for using spectral measurements for ranging in LIDAR and sensing applications. 
\end{abstract}

%\setboolean{displaycopyright}{true}

\maketitle

Space-time wave packets (STWPs) are a class of spatiotemporally structured pulsed optical beams that have attracted considerable interest over the past decade \cite{Yessenov22AOP}. The hallmark of STWPs is the tight association between the spatial frequencies undergirding the beam transverse spatial profile and the temporal frequencies (or wavelengths) undergirding the pulse longitudinal temporal profile \cite{Kondakci16OE,Parker16OE,Kondakci17NP,Yessenov19PRA}. STWPs have roots extending back four decades to Brittingham's discovery of the focus-wave mode \cite{Brittingham83JAP,Reivelt00JOSAA,Reivelt02PRE}, and subsequently X-waves \cite{Saari97PRL}, but with theoretical investigations outpacing experimental developments \cite{Turunen10PO,FigueroaBook14}.

The recent surge of interest in STWPs \cite{Porras17OL,Efremidis17OL,Wong17ACSP2,Wong21OE,Bejot21ACSP,Guo21Light,Guo21PRR,Shiri22Optica,Ramsey23PRA,Stefanska23ACSP} stems from the ease of synthesizing such unique field configurations with high precision \cite{Yessenov19OE}, which enables quantitative comparison between theoretical predictions and measurements over a surprisingly broad span of values in the STWP parameter space. The unique spatiotemporal spectral signature of STWPs leads to a host of useful characteristics, including propagation invariance in linear media \cite{Bhaduri19OL} (even in presence of group-velocity dispersion \cite{Malaguti08OL,Malaguti09PRA,He22LPR,Hall23NatPhys,Hall23LPR}), self-healing \cite{Kondakci18OL}, tunable group velocity \cite{Salo01JOA,Kondakci19NC}, anomalous refraction \cite{Bhaduri20NP}, among other intriguing features \cite{Diouf22SR,Diouf23Optica}.

A crucial feature of STWPs has been recently confirmed \cite{Hall23arxiv}: their predicted \cite{Bhaduri18OE} propagation invariance over large distances at the kilometer range. This demonstration thus opens the way to verifying the unique features of STWPs over large distance in real-world settings. This experiment utilizing a broadband pulsed beam can be compared to the handful of kilometer-scale tests of monochromatic diffraction-free beams; e.g., Bessel beams \cite{Aruga99AO} and abruptly focusing beams \cite{Zhang19APLP}.

A particularly intriguing feature of STWPs is the possibility of varying a selected parameter axially, while holding its other features invariant; e.g., accelerating STWPs \cite{Clerici08OE,ValtnaLukner09OE,Li20SR,Yessenov2020PRLaccel,Li21CP,Hall22OLArbAccel} (axially varying group velocity), and axial spectral encoding \cite{Motz21PRA}, in which the on-axis spectrum varies almost arbitrarily along the propagation axis; e.g., red-shifting, blue-shifting spectra, or compounded axial spectral variation. The proof-of-principle experiment in \cite{Motz21PRA} was conducted over short propagation distances in a well-controlled laboratory environment.  Validating the potential for using spectrally encoded STWPs in target ranging by correlating its axial location with the scattered wavelength requires testing its performance over extended distances in open air.

Here we demonstrate axial spectral encoding over a distance of 200~m in an open-air laser range, and the synthesized STWPs are directed down the range where spectral measurements are performed. Using a 25-nm-bandwidth source, the on-axis bandwidth for the spectrally encoded STWPs is $\approx\!5$~nm, and the central wavelength varies with axial propagation -- across the entire source bandwidth -- in both red-shifting and blue-shifting configurations. We also demonstrate an alternate configuration in which the on-axis spectrum is spectrally shifted without recourse to spectral filtering by changing an internal degree of freedom of the STWP. These results suggest potential applications in LIDAR and sensing, where axial ranging may be carried out directly through spectral measurements. 

\begin{figure}[t!]
\centering
\includegraphics[width=8.6cm]{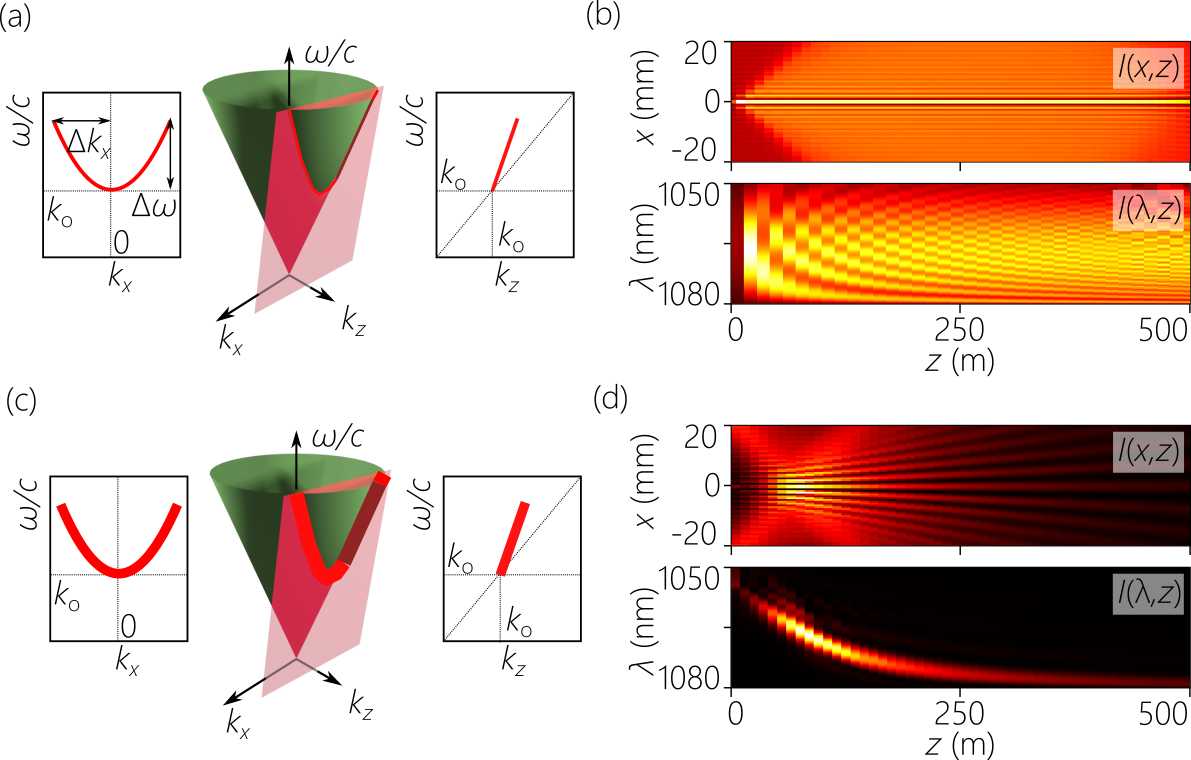}
\caption{Concept of axial spectral encoding. (a) Spectral support of a propagation-invariant STWP at the intersection of the light-cone surface $k_{x}^{2}+k_{z}^{2}\!=\!(\tfrac{\omega}{c})^{2}$ with a tilted spectral plane, along with its projections onto the $(k_{x},\tfrac{\omega}{c})$ and $(k_{z},\tfrac{\omega}{c})$ planes. (b) The axial evolution of the time-averaged intensity $I(x,z)$, and the on-axis ($x\!=\!0$) spectrum $I(\lambda,z)$. (c-d) Same as (a-b) for an STWP endowed with red-shifting axial spectral encoding.}
\label{Fig:Concept}
\end{figure}

We first summarize the theoretical foundation for axial spectral encoding with STWPs [Fig.~\ref{Fig:Concept}]. A particularly useful visualization tool is the spectral support on the surface of the free-space light-cone $k_{x}^{2}+k_{z}^{2}\!=\!(\tfrac{\omega}{c})^{2}$ \cite{Donnelly93PRSLA}; here, $k_{x}$ is the transverse wave number along $x$, $k_{z}$ is the axial wave number along the propagation axis $z$, we take the field (without loss of generality) to be uniform along the other transverse dimension $y$ ($k_{y}\!=\!0$), $\omega$ is the temporal frequency, and $c$ is the speed of light in vacuum. The spectral support of a propagation-invariant STWP is the conic section at the intersection of the light-cone with a tilted plane $\omega-\omega_{\mathrm{o}}\!=\!(k_{z}-k_{\mathrm{o}})c\tan\theta$, which is parallel to the $k_{x}$-axis and makes an angle $\theta$ (the spectral tilt angle) with the $k_{z}$-axis, $\omega_{\mathrm{o}}$ is a carrier frequency, and $k_{\mathrm{o}}\!=\!\tfrac{\omega_{\mathrm{o}}}{c}$ [Fig.~\ref{Fig:Concept}(a)]. In the narrowband paraxial regime, the spectral support in the vicinity of $k_{x}\!=\!0$ is approximated by a parabola, $\tfrac{\omega-\omega_{\mathrm{o}}}{\omega_{\mathrm{o}}}\!=\!\tfrac{k_{x}^{2}}{2k_{\mathrm{o}}^{2}(1-\cot\theta)}$. The STWP takes the form $E(x,z;t)\!=\!e^{i(k_{\mathrm{o}}z-\omega_{\mathrm{o}}t)}\psi(x,z;t)$, where $\psi(x,z;t)\!=\!\psi(x,0;t-z/\widetilde{v})$ is a slowly varying envelope that travels rigidly without diffraction or dispersion at a group velocity $\widetilde{v}\!=\!c\tan\theta$ \cite{Kondakci19NC,Yessenov19OE}. The time-averaged intensity $I(x,z)\!=\!\int\!dt\,|\psi(x,z;t)|^{2}$ is thus diffraction-free, $I(x,z)\!=\!I(x,0)$.

The transverse intensity profile $I(x,z)$ takes here the form of a narrow central feature of width $\Delta x$ atop a pedestal (this pedestal does not exist when the STWP is localized in both transverse dimensions \cite{Guo21Light,Yessenov22NC}). The width $\Delta x$ is proportional to the inverse of the spatial bandwidth $\Delta k_{x}$, which is in turn determined by the temporal bandwidth $\Delta\omega$ and the spectral tilt angle $\theta$. We plot in Fig.~\ref{Fig:Concept}(b) the axial evolution of the intensity $I(x,z)$ and the on-axis ($x\!=\!0$) spectrum, which is stable and corresponds to the full STWP spectrum. The spectral oscillations observed are due to the finite system aperture used in the calculation, which is taken from the experimental configuration (see below).

In practice, however, no finite system can produce an ideal association between spatial and temporal frequencies, $k_{x}$ and $\omega$, respectively, which implies infinite energy and propagation distance \cite{Sezginer85JAP}. Instead, there is always a spectral uncertainty $\delta\omega$ in this association for a finite-energy wave packet, which determines its propagation distance $L_{\mathrm{max}}\!\sim\!\tfrac{\delta\omega}{c|1-\cot{\theta}|}$, so that increasing $L_{\mathrm{max}}$ necessitates reducing $\delta\omega$ and $\delta\theta$, where $\theta\!=\!45^{\circ}+\delta\theta$. This formula is independent of the system aperture for a wide range of values of $\theta$ and $\delta\omega$, as validated in \cite{Yessenov19OE}. However, as we recently showed in \cite{Hall23arxiv}, when $\delta\theta\!\rightarrow\!0$, that this formula for $L_{\mathrm{max}}$ no longer holds, and instead we have $L_{\mathrm{max}}\!\sim\!\tfrac{W\Delta x}{\lambda_{\mathrm{o}}}$, where $W$ is the aperture width. This is the regime in which the measurements reported here are performed.

To produce an STWP whose on-axis spectrum is no longer invariant requires changing its spectral support on the light-cone [Fig.~\ref{Fig:Concept}(c)]. Associating each temporal frequency in the STWP with a prescribed spatial spectrum, rather than a single spatial frequency, reduces $L_{\mathrm{max}}$, but the STWP can now be endowed with a controllable on-axis spectrum [Fig.~\ref{Fig:Concept}(d)]. Rather than the full spectrum being recorded on-axis along $z$, only a narrow spectrum is recorded, and for the example in Fig.~\ref{Fig:Concept}(d) this spectrum red-shifts along $z$ (i.e., the central wavelength of the narrow spectrum shifts to longer values with $z$).

\begin{figure}[t!]
\centering
\includegraphics[width=8.4cm]{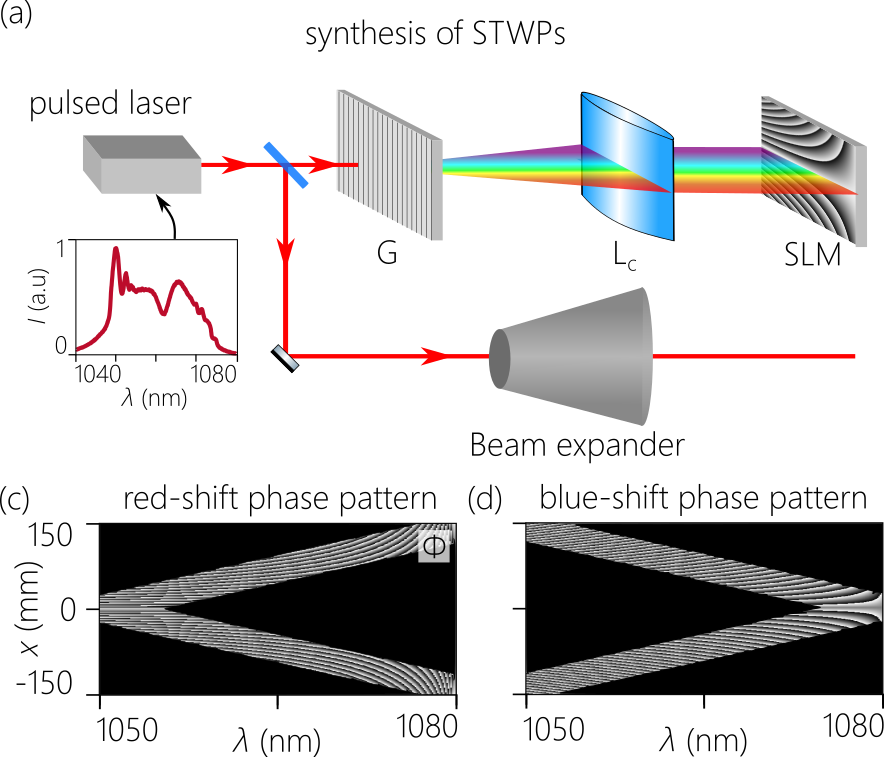}
\caption{(a) Schematic of the STWP synthesis setup. G: Diffraction grating; L$_{\mathrm{c}}$: cylindrical lens; SLM: spatial light modulator. The SLM phase pattern displayed is that for a propagation-invariant STWP. (b,c) Plots of the SLM phase to produce axially encoded STWPs by amplitude filtering of the propagation-invariant STWP shown in (a).}
\label{Fig:Setup}
\end{figure}

The STWPs are synthesized via the setup in Fig.~\ref{Fig:Setup}(a), using pulses from a mode-locked fiber laser (Spark Laser, Alcor) of width $\Delta T\!\approx\!100$~fs, bandwidth $\Delta\lambda\!\approx\!25$~nm, and average power $\sim\!2$~W. The laser beam is magnified $15\times$ to a width of $\approx\!25$~mm, and directed to a diffraction grating (1200~lines/mm, $50\!\times\!50$~mm$^2$). The -1~diffraction order is collimated with a cylindrical lens (focal length $f\!=\!400$~mm) onto a spatial light modulator (SLM; Meadowlark, E19X12) that imparts a 2D phase distribution $\Phi$ to the spectrally resolved wave front. In the case of a propagation-invariant STWP, we associate a spatial frequency $k_{x}(\lambda)$ to each wavelength $\lambda$, $\Phi(x,\lambda)\!=\!\pm k_{x}(\lambda)x$ [Fig.~\ref{Fig:Setup}(a)]. The retro-reflected wave front returns from the SLM through the lens to the grating, whereupon the STWP is constituted, and is then magnified to a final aperture of $\approx\!250$~mm using a telescope consisting of a singlet lens ($f\!=\!150$~mm, 25-mm aperture) and a large parabolic mirror ($f\!=\!3$~m, 300-mm aperture).

To produce the spectrally encoded STWPs, the SLM phase distribution $\Phi$ is modified by adding amplitude masking to the spectrally resolved wave front at the SLM. Two examples are shown in Fig.~\ref{Fig:Setup}(b,c) corresponding to red-shifting and blue-shifting spectrally encoded STWPs. Both of these start from the phase $\Phi$ for the propagation-invariant STWP at a particular $\theta$ [Fig.~\ref{Fig:Setup}(a)], after which an amplitude mask is added \cite{Motz21PRA}. The complex SLM reflectance is $r(\lambda,x)\!=\!\pm k_{x}(\lambda)x\cdot\mathrm{rect}\left(\tfrac{x\pm x_{\mathrm{o}}(\lambda)}{W_{\mathrm{o}}(\lambda)}\right)$, where $\mathrm{rect}(\tfrac{x}{W})$ is a flat-top function of width $W$, the wavelength-dependent shifted center $x_{\mathrm{o}}(\lambda)$ determines the axial distribution of the on-axis spectrum $I(\lambda,z)$ through $z(\lambda_{\mathrm{c}})\!\sim\!\tfrac{k}{k_{x}}x_{\mathrm{o}}(\lambda_{\mathrm{c}})$, where $\lambda_{\mathrm{c}}$ is the center of the on-axis spectrum, and the wavelength-dependent width $W_{\mathrm{o}}(\lambda)$ determines the transverse width over which the spectrum is spread.

\begin{figure}[t!]
\centering
\includegraphics[width=8cm]{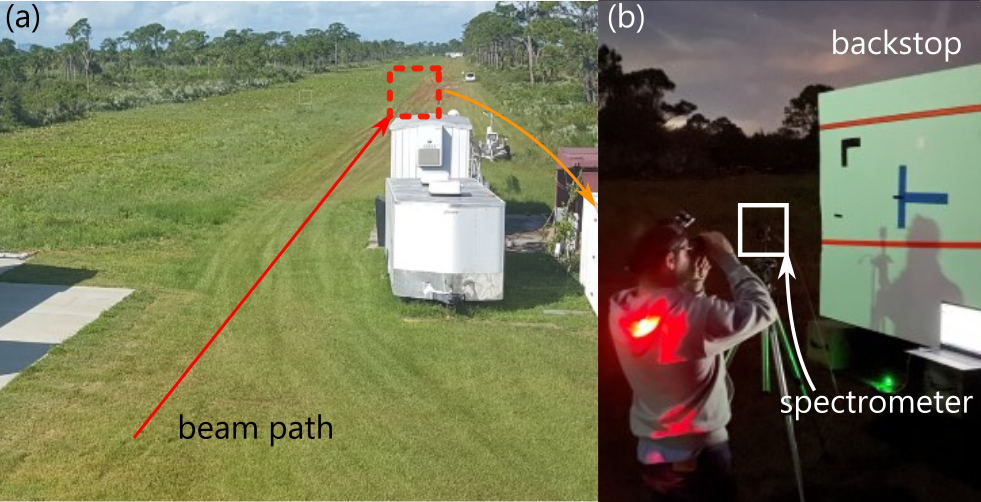}
\caption{The measurement configuration at TISTEF. (a) Photograph of the outdoor range. The solid red arrow indicates the overall beam path down the range. (b) Photograph of the measurement apparatus moving down the range.}
\label{Fig:TISTEF}
\end{figure}

We carried out our experiments at the Townes Institute Science and Technology Experimentation Facility (TISTEF), which is a 1-km terrestrial laser range on the Florida space coast. The setup in Fig.~\ref{Fig:Setup}(a) is housed in a mobile laboratory, the synthesized STWP exits it through a window, and a flat mirror (300-mm aperture) directs the beam path to a measurement station placed on a golf cart that is driven down the range [Fig.~\ref{Fig:TISTEF}(a)]. A backstop intercepts the STWP; a SWIR camera (Golden Eye, CL-033) attached to a telescope that is placed in the mobile laboratory images the backstop to help guide the direction of the STWP via the flat mirror; and a range-finder (Burris, 300351) is used to determine the axial location $z$ of the backstop (and thus the propagation distance) down the range [Fig.~\ref{Fig:TISTEF}(b)]. Once the STWP is located, a multimode optical fiber (diameter 200 $\mu$m) held on a tripod is aligned to intercept its main central lobe. The fiber is connected to a spectrometer (ThorLabs CCS175, spectral resolution $\delta\lambda\!\approx\!1$~nm) to measure the on-axis spectrum. Using this configuration, measurements were carried out over an axial range $z\!\sim\!100-200$~m from the laboratory window down the laser range. To minimize the impact of extraneous light and any effects of jitter resulting from turbulence, the measurements were taken in the late evening (after 9~pm local time) with a $C_{n}^{2}<10^{-14}$~m$^{-2/3}$.

\begin{figure}[t!]
\centering
\includegraphics[width=8.6cm]{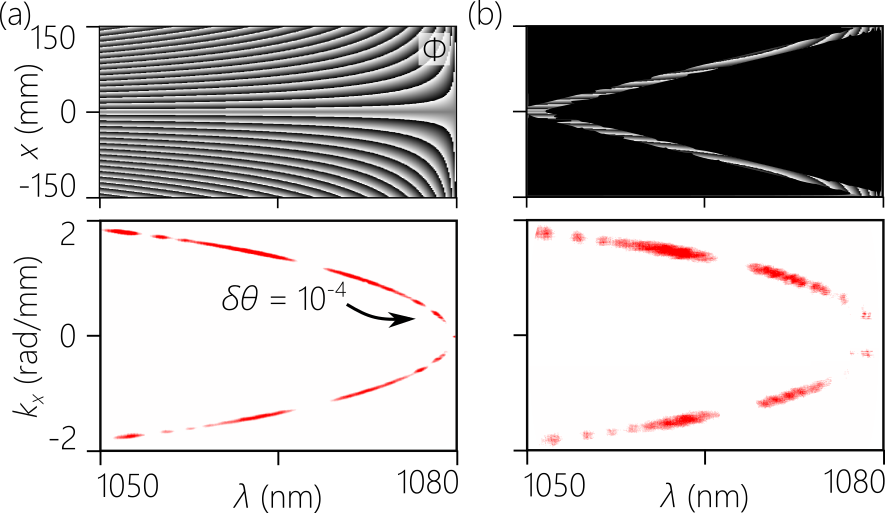}
\caption{The phase patterns $\Phi(\lambda,x)$ accompanied by the measured spatiotemporal spectra in $(k_{x},\lambda)$-space for the (a) propagation-invariant STWP with $\delta\theta\!=\!10^{-4}$ (in degrees) and (b) the red-shifted axially varying spectrum.}
\label{Fig:Spectral}
\end{figure}

At each axial position $z$, the phase distributions are cycled between three alternatives. First, a propagation-invariant STWP with $\delta\theta\!\approx\!10^{-4}$ is synthesized. The SLM phase pattern and the measured spatiotemporal spectrum for this STWP are plotted in Fig.~\ref{Fig:Spectral}(a). The on-axis spectrum for such an STWP is expected at any $z$ to be the same as that of the source, with minimal spectral changes up to $L_{\mathrm{max}}$. Second, a red-shifting spectrally encoded STWP is synthesized while holding $\delta\theta\!\approx\!10^{-4}$ fixed by amplitude masking of the SLM phase distribution associated with the propagation-invariant STWP. The SLM phase and the measured spatiotemporal spectrum are plotted in Fig.~\ref{Fig:Spectral}(b); note the broadening of the parabolic spectrum compared to Fig.~\ref{Fig:Spectral}(a). We expect the on-axis spectrum at any $z$ to be narrow, and for its central wavelength $\lambda_{\mathrm{c}}$ to shift from short to long wavelengths along $z$. Third, we produced a blue-shifting spectrally encoded STWP, that is similar to its red-shifting counterpart, except $\lambda_{\mathrm{c}}$ is expected to shift from long to short wavelengths along $z$. 

\begin{figure}[t!]
\centering
\includegraphics[width=8.6cm]{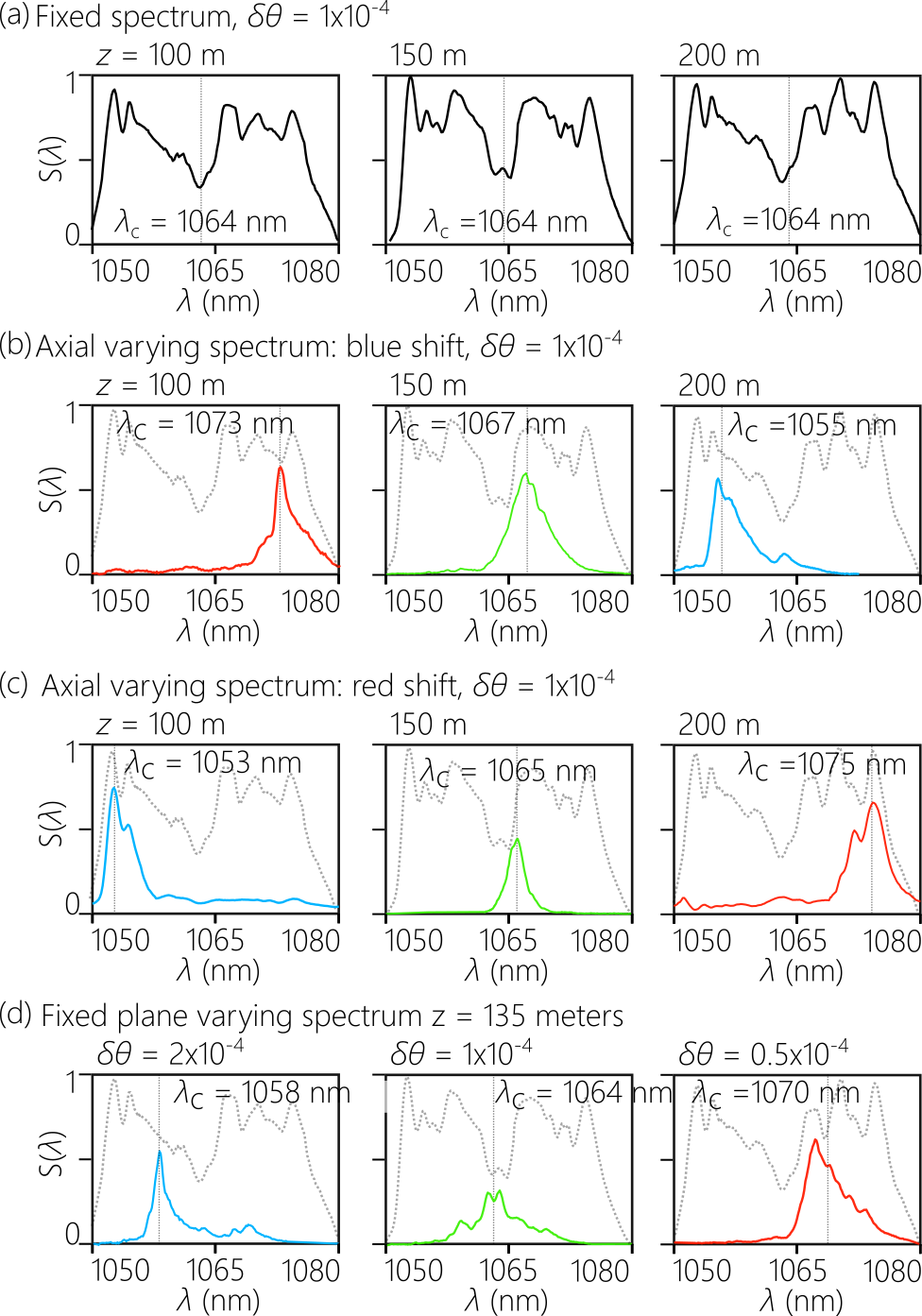}
\caption{Measured axially varying spectra at $z\!=\!100,150$, and 200~m for (a) a propagation-invariant STWP with $\delta\theta\!=\!\times10^{-4}$, (b) the red-shifting spectrally encoded STWP with $\lambda_{\mathrm{c}}\!=\!1053,1065$, and 1075~nm, and (c) the blue-shifting STWP with $\lambda_{\mathrm{c}}\!=\!1073,1067$, and 1055~nm. (d) Measured spectrum at a fixed location $z\!=\!135$~m while varying the spectral tilt angle $\delta\theta\!=\!2\times10^{-4},1\times10^{-4}$, and $5\times10^{-5}$, corresponding to $\lambda_{\mathrm{c}}\!=\!1058,1064$, and 1070~nm. Values of $\delta\theta$ are in degrees.}
\label{Fig:Measurements}
\end{figure}

Measurements of the on-axis spectrum for the propagation-invariant STWP are plotted in Fig.~\ref{Fig:Measurements}(a) for $z\!=\!100,150$, and 200~m. The measured spectrum is stable along this propagation distance, and it matches the spectrum of the initial laser [Fig.~\ref{Fig:Setup}(a)]. The central wavelength is $1064$~nm, and the spectrum extends over a bandwidth $\Delta\lambda\!\approx\!25$~nm. The measured on-axis spectra for the alternative scenario of a red-shifting spectrally encoded STWP are plotted in Fig.~\ref{Fig:Measurements}(b). Here the on-axis spectrum has a FWHM bandwidth of $\Delta\lambda\!\approx\!5$~nm (rather than the full source spectrum $\Delta\lambda\!\approx\!25$~nm), and a central wavelength at $\lambda_{\mathrm{c}}\!=\!1053,1065$, and $1075$~nm recorded at $z\!=\!100,150$, and 200~m, respectively. The measured on-axis spectra for the blue-shifting spectrally encoded STWP are plotted in Fig.~\ref{Fig:Measurements}(c). The spectra for this blue-shifting SWTP are similar to those for its red-shifted counterpart, except that the spectra occur in reverse order along the axis, with $\lambda_{\mathrm{c}}\!=\!1073,1067$, and $1055$~nm at $z\!=\!100,150$, and 200~m, respectively. The change in $\lambda_{\mathrm{c}}$ is approximately linear along $z$ as predicted from the implemented SLM phase. Note that an even narrower on-axis spectrum can be readily produced by modifying the SLM phase in Fig.~\ref{Fig:Spectral}(b). Finally, we verify that the spectrum can be shifted at a fixed axial location (here at $z\!\approx\!135$~m) by varying the spectral tilt angle, $\delta\theta\!=\!2\times10^{-4},\times10^{-4}$, and $0.5\times10^{-4}$, resulting in a spectral shift $\lambda_{\mathrm{c}}\!=\!1058,1064$, and $1070$~nm [Fig.~\ref{Fig:Measurements}(d)].

In contrast to our previous long-range propagation experiments \cite{Bhaduri19OL,Hall23arxiv}, the STWPs are synthesized here using an SLM rather than a high-resolution phase plate \cite{Kondakci18OE}. This indicates the surprisingly utility of SLMs in ultra-precise spatiotemporal modulation of STWPs. Future work will focus on collecting the light scattered from objects placed in the path of spectrally encoded STWPs, with the aim of verifying the efficacy of direct spectral measurements in axial ranging. Finally, although we made use of STWPs in the form of light sheets (the field is localized along only one spatial dimension), recent progress has enabled synthesizing STWPs that are localized along all dimensions \cite{Guo21Light,Yessenov22NC}, which have yet to be tested at the kilometer scale.

%\clearpage
In conclusion, we have demonstrated axial spectral encoding using STWPs in an open-air laser range over a distance of 200~m. When using propagation-invariant STWPs, the full spectrum is present on axis and is stable with propagation. With axial spectral encoding, the same total source spectrum leads to an on-axis spectrum comprising only a portion of that spectrum, and the STWP can be designed for this spectrum to undergo prescribed axial dynamics; e.g., blue-shifting or red-shifting with propagation distance. We have also verified another configuration where the on-axis spectral segment is tuned across the available spectrum without spectral filtering, but rather by tuning an internal degree of freedom of the STWP (its spectral tilt angle). These results suggest potential applications in LIDAR and remote sensing, where axial ranging information may be captured through purely spectral measurements.

\section*{Funding}
U.S. Office of Naval Research (ONR) N00014-19-1-2192 and N00014-20-1-2789.

\section*{Data availability}
Data underlying the results presented in this paper are not publicly available at this time but may be obtained from the authors upon reasonable request.

\section*{Disclosures}
The authors declare no conflicts of interest.

\section*{Acknowledgements}
We thank the TISTEF staff for their support.

\section*{Data availability}
Data underlying the results presented in this paper are not publicly available at this time but may be obtained from the authors upon reasonable request.

% Bibliography
\bibliography{diffraction}

\end{document}